
%
%
\input harvmac
\def\ie{{\it i.e.}}

\def\no{\noindent}
\def\o{\over}
\def\nl{\hfill\break}
\def\alp{\alpha}\def\gam{\gamma}\def\del{\delta}\def\lam{\lambda}
\def\eps{\epsilon}
\def\sig{\sigma}\def\th{\theta}\def\om{\omega}
\def\thb{{\th\kern-0.465em \th}}
\def\ZZ{{\bf Z}}

\def\do{\downarrow}
\def\up{\uparrow}
\def\sm{$S$-matrix}
\def\Alm{ A_\ell^{(m)}} \def\Blm{ B_\ell^{(m)}}
\def\qed{\hfill$\vrule height 2.5mm width 2.5mm depth 0mm$}


\nref\rGH{M.C. Gutzwiller, Phys. Rev. Lett. 10 (1963) 159; \nl
 J. Hubbard, Proc. Roy. Soc. (London), Ser. A, 276 (1963) 238.}
\nref\rAnd{P.W. Anderson, Science 235 (1987) 1196.}
\nref\rLiebWu{E.H. Lieb and  F.Y. Wu, Phys. Rev. Lett. 20 (1968) 1445.}
\nref\rEK{F.H.L. E\ss ler and V.E. Korepin, Phys. Rev. Lett. 72 (1994)
 908, and Stony Brook preprint
 ITP-SB-93-45, cond-mat/9310\-056, to appear in Nucl. Phys. B.}
\nref\rMW{P.K. Mitter and P.H. Weisz, Phys. Rev. D8 (1973) 4410.}
\nref\rGN{D. Gross and A. Neveu, Phys. Rev. D10 (1974) 3235.}
\nref\rDF{R. Dashen and Y. Frishman, Phys. Rev. D11 (1975) 2781.}
\nref\rBHN{T. Banks, D. Horn and H. Neuberger, Nucl. Phys. B108
 (1976) 119.}
\nref\rBKWK{B. Berg, M. Karowski, P. Weisz and V. Kurak, Nucl. Phys.
  B134 (1978) 125.}
\nref\rBW{B. Berg and P. Weisz, Nucl. Phys. B146 (1978) 205.}
\nref\rBel{A.A. Belavin, Phys. Lett. B87 (1979) 117.}
\nref\rAL{N. Andrei and J.H. Lowenstein, Phys. Rev. Lett. 43 (1979) 1698,
  and Phys. Lett. B91 (1980) 401.}
\nref\rDL{C. Destri and J.H. Lowenstein, Nucl. Phys. B205 (1982) 369.}
\nref\rSmir{F.A.~Smirnov, {\it Form Factors in Completely Integrable
 Models of Quantum Field Theory} (World Scientific, Singapore, 1992)}
\nref\raff{I. Affleck, talk given at the Nato Advanced
 Study Institute on {\it Physics, Geometry and Topology}, Banff, August
 1989.}
\nref\rLuk{S. Lukyanov, Rutgers preprint RU-93-30, hep-th/9307196.}
\nref\rher{T.R. Klassen and E. Melzer, Int. J. Mod. Phys. A8 (1993) 4131.}
\nref\rkink{T.R. Klassen and E. Melzer, Nucl. Phys. B382 (1992) 441.}
\nref\rMcW{B.M. McCoy and T.T. Wu, Phys. Lett. B87 (1979) 50.}
\nref\rnatan{N. Andrei, summer course on {\it Low-dimensional
 Quantum Field Theories for Condensed Matter Physicists}, Trieste 1992,
 unpublished.}
\nref\rWoy{F. Woynarovich, J. Phys. C16 (1983) 5293 and 6593.}
\nref\rGR{I.S. Gradshteyn and I.M. Rizhik, {\it Table of Integrals,
 Series, and Products} (Academic Press, Orlando, 1980).}
\nref\rFB{M.E. Fisher and M.N. Barber, Arch. Rational Mech. Annals
 47 (1972) 205.}
\nref\rOv{A.A. Ovchinnikov, Sov. Phys. JETP 30 (1970) 1160.}
\nref\rTaka{M. Takahashi, Progr. Theor. Phys. 43 (1970) 1619.}
\nref\raly{Al.B. Zamolodchikov, Nucl. Phys. B358 (1991) 619.}
\nref\rzamsb{A.B. Zamolodchikov and Al.B. Zamolodchikov, Nucl. Phys.
 B379 (1992) 602.}
\nref\rFS{P. Fendley and H. Saleur, preprint USC-93-022, hep-th/9310058.}
\nref\rzamsa{A.B.~Zamolodchikov and Al.B.~Zamolodchikov, Ann.~Phys.~120
 (1979) 253.}
\nref\rWeisz{P.H. Weisz, Nucl. Phys. B122 (1977) 1.}
\nref\rFT{L.D. Faddeev and L. Takhtajan, Phys. Lett. A85 (1981) 375, and
 J. Sov. Math. 24 (1984) 241.}
\nref\rBKW{B. Berg, M. Karowski and P. Weisz, Phys. Rev. D19 (1979) 2477.}
\nref\rMTW{ T.T. Wu, B.M. McCoy, C.A. Tracy and E. Barouch, Phys. Rev.
 B13 (1976) 316; ~B.M. McCoy, C.A. Tracy and T.T. Wu, Phys. Rev. Lett. 38
 (1977) 793.}
\nref\rLash{M.Yu. Lashkevich, preprint LANDAU-94-TMP-4, hep-th/9406118.}

\Title{\vbox{\baselineskip12pt\hbox{TAUP 2203-94}\hbox{cond-mat/9410043} }}
{\vbox{\centerline{On the Scaling Limit of the} \vskip13pt
       \centerline{1D Hubbard Model at Half Filling}}}
\medskip\centerline{Ezer Melzer}
\medskip\centerline{\it School of Physics and Astronomy}
\smallskip\centerline{\it Beverly and Raymond Sackler Faculty
  of Exact Sciences}
\smallskip\centerline{\it Tel-Aviv University}
\smallskip\centerline{\it Tel-Aviv 69978, ISRAEL}
\medskip\centerline{email: melzer@ccsg.tau.ac.il}
\vskip 13mm

\centerline{{\bf Abstract}}
\vskip 3mm

The dispersion relations and \sm\ of the one-dimensional Hubbard
model at half filling are considered in a certain scaling limit.
(In the process we derive a useful small-coupling expansion of the
exact lattice dispersion relations.)
The resulting scattering theory is consistently identified
as that of the $SU(2)$ chiral-invariant Thirring
(or Gross-Neveu) field theory,
containing both massive and massless sectors.

\Date{\hfill}
\vfill\eject

\newsec{Introduction}
\ftno=0

The Hubbard model~\rGH~describes electrons on a lattice with
on-site interaction only, in addition to a standard
nearest-neighbor hopping term. In two dimensions the model
has received much attention lately in connection with high-$T_c$
superconductivity.
Some of its properties are believed~\rAnd~to be similar to those
exhibited by the one-dimensional model, which is exactly
solvable by means of the Bethe Ansatz technique~\rLiebWu.
In this paper we  discuss certain aspects of the
scaling limit of the one-dimensional model, which are
relevant for its large-distance asymptotic behavior.

The hamiltonian of the linear Hubbard model is given by
\eqn\ham{ H ~=~ -{1\o 2} \sum_{j=1}^L \sum_{\sig=\up,\do}
 \left(c^\dagger_{j,\sig} c_{j+1,\sig}+c^\dagger_{j+1,\sig} c_{j,\sig}\right)-
  2U\sum_{j=1}^L (n_{j,\up}-{1\over 2})(n_{j,\do}-{1\over 2})~,}
where $c_{j,\sig}$ are canonical fermionic annihilation operators,
$j$ labels the sites of a periodic
chain of length $L$ (which is taken to be even), $\sig$ labels the
two spin degrees of freedom, $U$ is a real coupling constant, and
$n_{j,\sig}=c^\dagger_{j,\sig} c_{j,\sig}$ is the number operator for spin
$\sig$ on site $j$.
(The overall normalization of $H$ chosen in \ham\ will be convenient
later on.)
Since $H$ commutes with the total number operator $\sum_{j=1}^L
\sum_{\sig=\up,\do} n_{j,\sig}$, it can be diagonalized separately
in eigenspaces of fixed number of ``electrons'' $N$.

The model has\foot{For details on the following features of the
model see~\rEK~and references therein.}
 an $SO(4)=SU(2)\times SU(2)/\ZZ_2$ symmetry, the
two $SU(2)$'s pertaining to spin ($s$) and charge ($c$).
The spectrum is built out of four fundamental excitations
(alias quasiparticles),
forming a ``spinon-antispinon'' $SU(2)_s$-doublet
and a ``holon-antiholon'' $SU(2)_c$-doublet.
However, this separation of spin and charge seen in the quantum
numbers of the quasiparticles does not mean that the theory decouples
into a tensor product of two $SU(2)$-symmetric models.
For instance, there is a selection rule which allows only representations
with integer total $SU(2)_s$ {\it and} $SU(2)_c$ spin in the spectrum
(implementing the $\ZZ_2$ quotient which reduces the symmetry from
the naive $SU(2)\times SU(2)$ down to $SO(4)$).\foot{In particular,
this restriction implies that there are no single-particle states
in the spectrum and thus the fundamental quasiparticles are ``confined''.}

The situation at half-filling $N=L$ is of special interest. In this
case one of the two quasiparticle doublets develops a mass gap while the
other remains massless. The fact that the mass gap vanishes as the
coupling tends to zero opens up the possibility for the existence
of a scaling limit in which both massive and massless excitations
survive in the spectrum of the resulting field theory. Our aim is to
explore this possibility, which we will do at the level of
the dispersion relations of the quasiparticles and their scattering
amplitudes. The \sm\ theory obtained this way is then identified
as that of the $SU(2)$
chiral-invariant Thirring (or Gross-Neveu) model, whose lagrangian
is given by~[5-7]        
\eqn\lagr{ \eqalign{ \cal{L} ~&=~ i\bar{\psi}\rlap/\partial \psi
 +g\bigl[ (\bar{\psi}\psi)^2 -(\bar{\psi}\gam^5 \psi)^2 \bigr] \cr
                        &=~ i\bar{\psi}\rlap/\partial \psi
 -{1\o 2}g\bigl[ (\bar{\psi}\gam^\mu \psi)^2
  +\sum_{a=1}^3 (\bar{\psi}\gam^\mu \sig^a \psi)^2 \bigr]~~,\cr}}
where $\psi$ is a doublet of Dirac spinors, $\gam^\mu$ are
Dirac matrices in 1+1 dimensions, and $\sig^a$ are the Pauli matrices
(the equality between the two lines of \lagr\ can be established
with the aid of identities listed in the appendix of~\rMW).

The field theory \lagr\ and its \sm\ have been discussed
in [8-16].    
It is known,
using bosonization, that the theory essentially decouples
into a massless and a massive sector.
(This statement holds modulo certain orbifolding, cf.~\raff\rher,
which is reflected for instance by ``kinky'' restrictions~\rkink~on
the multiparticle spectrum.)
The  massless sector is described
by the level one $SU(2)$ WZW conformal field theory. The massive
sector, on the other hand, can be viewed as a marginally
relevant $SU(2)$-preserving (integrable)
perturbation of another copy of the same conformal
field theory, where mass is generated dynamically through
``dimensional transmutation'' (for a construction of this
sector of the theory from a scaling limit of the XXZ spin chain
cf.~\rMcW).\foot{Somewhat confusingly,
in the literature the \sm\ of the massive
sector alone is occasionally referred to as that of the full theory \lagr.}

The emergence
of the $SU(2)$ chiral-invariant Thirring field theory
from the scaling limit of the half-filled Hubbard model,
as described in the sequel, is not surprising.
It was already noted on the basis of renormalization group
and symmetry arguments in~\raff, where the continuum (low-energy)
limit of \ham\ was considered.
Nevertheless, we think that our complementary analysis
is worthwhile.
Related work can be found in~\rnatan.

The rest of the paper is organized as follows.
In sect.~2 we define the scaling limit and derive the scaled
dispersion relations. In the process
the familiar Hubbard model dispersion relations (2.1)--(2.2) are rewritten
in the form (2.3)--(2.4), which is most useful for analyzing the
rather singular zero-coupling limit; we find this apparently
new form, whose derivation is presented in the appendix,
interesting  by itself. The scaled \sm\ obtained in sect.~3
is discussed in the final section.

\newsec{The scaling limit}

We restrict attention to the attractive regime $U>0$, where the spin
excitations (spinons) are massive while the charge excitations (holons) are
massless. The repulsive regime is dual to the attractive one
in the sense that the properties of the two excitations are
interchanged~\rEK\rWoy.
The spin-wave dispersion relation is given in the parametric form
(see~\rEK \foot{Our conventions differ from those in~\rEK~by an overall
factor of ${1\o 2}$ in the hamiltonian \ham\ and a change in the sign
of $U$.} and references therein)
\eqn\sdisp{  \eqalign{
  p_s(k) ~&=~ k-\int_0^\infty {d\om \o \om}
   {J_0(\om) \sin(\om \sin k)\o \cosh(\om U)} e^{-\om U} \cr
  \eps_s(k) ~&=~ U-\cos k+\int_0^\infty {d\om \o \om}
   {J_1(\om) \cos(\om \sin k)\o \cosh(\om U)} e^{-\om U} ~~,\cr}}
where the $J_\nu(\om)$ are  Bessel functions.
The charge-wave dispersion relation reads
\eqn\cdisp{
  p_c(\lam) ~=~ -\int_0^\infty {d\om \o \om}
   {J_0(\om) \sin(\om \lam)\o \cosh(\om U)}  ~~~,~~~~~~
  \eps_c(\lam) ~=~ \int_0^\infty {d\om \o \om}
   {J_1(\om) \cos(\om \lam)\o \cosh(\om U)}~~,}
where, for nonzero $U$, $\lam\in (-\infty,\infty)$.
(In the free case $U$=0 one has
$|\lam|\leq 1$, and, using formulas 6.693(1-2) of~\rGR,
eq.~\cdisp\ reduces to ~$p_c(\lam)\bigl|_{U=0}=-\arcsin\lam$
{}~and~ $\eps_c(\lam)\bigl|_{U=0}=\sqrt{1-\lam^2}$, so that
{}~$\eps_c(p)\bigl|_{U=0}=\cos p$.)

In order to take the scaling limit we need a more convenient form
for the dispersion relations.
Let ~$U_n\equiv (n+{1\o 2}){\pi\o U}$. Then, as shown in
the appendix,
eq.~\sdisp\ can be expanded as
\eqn\sexp{ \eqalign{
 p_s(k) ~&=~ {2\o U} \sum_{n=0}^\infty {1\o U_n} K_0(U_n)
   \sinh(U_n \sin k) \cr
 \eps_s(k) ~&=~ {2\o U} \sum_{n=0}^\infty {1\o U_n} K_1(U_n)
   \cosh(U_n \sin k)~, \cr}}
for  ~$k\in (-{\pi \o 2},{\pi\o 2})$, while for $|\lam|\geq 1$
the following expansion of eq.~\cdisp\ is valid:
\eqn\cexp{ \eqalign{
 p_c(\lam) ~&=~ {\rm sgn}~\lam~\left( -{\pi\o 2} +{\pi\o U}
   \sum_{n=0}^\infty {(-1)^n\o U_n} I_0(U_n) e^{-U_n |\lam|}\right) \cr
 \eps_c(\lam) ~&=~ {\pi\o U} \sum_{n=0}^\infty {(-1)^n\o U_n} I_1(U_n)
       e^{-U_n |\lam|}~~. \cr}}
Here  $K_\nu(z)$ and $I_\nu(z)$ are the modified Bessel functions.

{}From eq.~\sexp\ we read off the spinon mass gap
(cf.~\rLiebWu\rFB)
\eqn\gap{ \Delta(U) ~=~ \eps_s(0) ~=~ {4\o \pi} \sum_{n=0}^\infty
  {K_1\bigl( (2n+1){\pi\o 2U} \bigr) \o 2n+1}~~,}
which vanishes for small $U$  according to~\rOv
\eqn\gaplead{ \Delta(U) ~\sim~ {4\sqrt{U}\o \pi}e^{-{\pi\o 2U}}
  ~~~~~~~{\rm as}~~U\to 0^+~.}
(For large $U$, on the other hand, the leading behavior is
$\Delta(U\gg 1) \sim {4\o \pi}\sum_{n=0}^\infty {U_n^{-1}\o 2n+1}={U\o 4}$;
in between $\Delta(U)$ increases monotonically for all $U>0$.)
The vanishing of the mass gap as $U\to 0$ is a necessary condition
for the existence in this limit of a nontrivial field theory
where the spinon survives as a massive particle of finite mass $M$.
Introducing a dimensionful lattice spacing $a$
(so that $H$ of \ham\ is replaced by $H/a$), we define
the corresponding {\bf scaling limit} by
\eqn\sclim{ a,U\to 0^+ ~~~~~~{\rm with}~~~~~~M \equiv
  \lim_{a,U\to 0} ~{\Delta(U)\o a} ~=   \lim_{a,U\to 0} ~
  {4\sqrt{U}\o \pi a}e^{-{\pi\o 2U}}~~~{\rm fixed}.}

To obtain a whole massive dispersion curve in the scaling limit,
the lattice rapidity variable $k$ has to be rescaled according to
\eqn\kresc{ k\to 0~~~~~~~~{\rm such~that}~~~~~~\th~=~{\pi \sin k\o 2U}
 ~~~~~~{\rm is~finite}.}
As a result, it follows that
only the $n$=0 term in \sexp\ survives in the limit \sclim, leading to
the scaled momentum and energy
\eqn\sEP{ P_s~=~\lim {p_s(k)\o a}~=~ M\sinh\th~~,~~~~~~~
          E_s~=~\lim {\eps_s(k)\o a}~=~ M\cosh\th~~.}
This is a relativistic massive dispersion relation
{}~$E_s(P)=\sqrt{M^2+P^2}$~ in 1+1 dimensions, parameterized in terms of
the customary rapidity variable $\th$ in the continuum. Note that the
overall normalization of the hamiltonian \ham\ is such that the ``speed
of light'' is 1.

Turning to the holon dispersion relation, we see from \cexp\ that
the energy $\eps_c(\lam)$ vanishes as $\lam\to\pm\infty$ (for fixed
$U$), the leading behavior being ~$\eps_c(|\lam| \gg 1)\sim
2I_1({\pi\o 2U})e^{-\pi |\lam|/2U}$. The momentum $p_c(\lam)$
approaches $\mp{\pi\o 2}$ in this limit, like
{}~$p_c(|\lam| \gg 1)\sim {\rm sgn}~\lam \bigl(-{\pi\o 2}+
2I_0({\pi\o 2U})e^{-\pi |\lam|/2U}\bigr)$.  This implies that
the dispersion curve linearizes around these points:
\eqn\clin{ \eps_c(p\sim \mp{\pi\o 2}) ~\sim~ v_c(U)({\pi\o 2}-|p|)~~~,
  ~~~~~~~v_c(U) ~=~{I_1({\pi\o 2U})\o  I_0({\pi\o 2U})}~~.}
The speed of the charge wave $v_c(U)$ was first given
in~\rOv~(cf.~also~\rTaka). We note that $v_c(U\to 0^+)=1$,
already suggesting that in the scaling limit the holons
become massless particles, traveling with the same ``speed of light''
as obtained from the massive dispersion relation of the spinons.

To obtain the full massless dispersion relation in the scaling
limit \sclim, we first note that as $U\to 0$ the holon
dispersion relation \cexp\ becomes linear for {\it all} $|\lam|\geq 1$,
and not just as $|\lam|\to\infty$. In fact, we may define
the rescaled rapidity variable $\beta$ by letting
\eqn\lresc{ \lam\to\pm 2 ~~~~~~{\rm such~that}~~~~\beta~=~
  \pm {\pi(2-|\lam|) \o 2U} ~~~~{\rm is~finite},}
when the limit \sclim\ is taken. Here and below the
upper and lower sign choices apply to right- and left-moving
excitations, respectively, which have to be treated separately.
 Now using the asymptotics of the functions $I_\nu(z)$, one
finds from \cexp\ and \lresc\ that the scaled dispersion relation
takes the form
\eqn\cEP{ P_c~=~\lim {1\o a} \left( p_c(\lam)\pm{\pi\o 2}\right)
               ~=~ \pm {M\o 2}e^{\pm\beta}~~,~~~~
          E_c~=~\lim { \eps_c(\lam)\o a}~=~ {M\o 2}e^{\pm\beta}~~,}
in the scaling limit \sclim.

Eq.~\cEP\ provides the standard
parameterization  [26-28]     
of the dispersion relation $E_c(P)=|P|$ of a massless particle in a
(1+1)-dimensional quantum field theory,
with $\beta\in(-\infty,\infty)$ (for both right- and
left-movers) and $M$ being some mass scale. This mass scale is
arbitrary and irrelevant if the theory is conformal, since no
observables depend on it and a change in it can be absorbed by
a redefinition of $\beta$. We find it satisfying that
in our case, where the theory does have a massive sector,
this mass scale turns out to be exactly equal to the mass
$M$ of the massive excitation \sEP, when one uses
the most natural  definition \lresc\ of $\beta$.
(Of course one can define $\beta$ as in \lresc\
but with $\lam\to\pm(2+{\rm const}\cdot U)$, which would have shifted
$\beta$ and hence rescaled $M$ of \cEP\ by a finite factor; the choice
const=0 is what we call the most natural one.)

\newsec{The \sm\ in the scaling limit}

In the previous section we found that the spectrum of the model
in the scaling limit \sclim\ consists of two doublets
(labeled by $s$ and $c$) of particles,
one  massive and the other massless. We can now use the
known \sm\ computed in~\rEK~for the spin chain to obtain the \sm\
of the continuum  theory, simply by reexpressing the amplitudes
of~\rEK~in terms of the rescaled variables \kresc, \lresc.

Due to the $SU(2)_s\times SU(2)_c$ symmetry of the model, the
two-particle \sm\
is block-diagonal with four  $4\times 4$ blocks
$S_{xy}$ ($x,y\in\{s,c\}$), corresponding to the
scattering sectors $s$-$s$, $s$-$c$, $c$-$s$, and $c$-$c$.
(In each block the rows and columns correspond to incoming and
outgoing particles; they are labeled by a pair of $SU(2)$
quantum numbers, namely spin up `$+$' or down `$-$'.)
In the spin chain, the amplitudes in each sector
are functions of a single variable $\mu$ which is defined as
\eqn\mudef{ \eqalign{
  \mu ~&=~ {|\sin k_1-\sin k_2|\o 2U}
   ~~~~~~~~~~~~~~~{\rm in}~~~s{\rm -}s \cr
 &=~{|\sin k-\lam|\o 2U}
   ~~~~~~~~~~~~~~~{\rm in}~~~s{\rm-}c~~({\rm or}~~c{\rm -}s) \cr
 &=~ {|\lam_1-\lam_2|\o 2U}
  ~~~~~~~~~~~~~~~~~~~~~~~{\rm in}~~~c{\rm -}c~, \cr}}
where the index $j$=1,2 refers to the two scattering particles.
In terms of the rescaled rapidities \kresc\ and \lresc,
where $U\to 0^+$, $k_j\to 0$, and $\lam_j\to \pm{\pi\o 2}$, this
becomes
\eqn\mudef{ \eqalign{
  \mu ~~&\to~~ {|\th_1-\th_2|\o \pi} ~~~~~~~~~~~~~~
     ~~~~~~~~~~~~~~~~~~~~~~~~~~{\rm in}~~~s{\rm -}s\cr
 &\to~~\infty ~~~~~~~~~~~~~~~~~~~~~~~~~~~
   ~~~~~~~~~~~~~~~~~~~~~~{\rm in}~~~s{\rm -}c \cr
 &\to~~ \cases{ -{|\beta_1-\beta_2|\o \pi}~~~&for~~R{-}R~~or~~L{-}L\cr
       \infty~~~&for~~R{-}L~~or~~L{-}R\cr}
     ~~~~~~~~{\rm in}~~~c{\rm -}c~, \cr}}
where R and L indicate right- and left-movers, respectively, in the
massless sector.

Using the results of~\rEK~we now find the following \sm\ amplitudes
in the scaling limit:
\eqn\Sss{ S_{ss}(\th_1,\th_2) ~=~ S_0(\th) \left(
  {\th \o \th-i\pi}I-{i\pi\o \th-i\pi}\Pi \right)~~,}
\eqn\Ssc{ S_{sc}(\th,\beta) ~=~ \lim_{\mu\to\infty}
  \left( -i~{1+ie^{\pi\mu}\o 1-ie^{\pi\mu}}~ I\right) ~=~ iI~,}
\eqn\Scc{ \eqalign{
  S_{cc}^{({\rm RR})}(\beta_1,\beta_2)~&=~
      S_{cc}^{({\rm LL})}(\beta_1,\beta_2)~=~-S_{ss}(\beta_1,\beta_2) \cr
  S_{cc}^{({\rm RL})}(\beta_1,\beta_2)~&=~\lim_{\mu\to\infty} S_0(-\mu)I
   ~=~iI~~,\cr}}
where $\th=\th_1-\th_2$, $I$ and $\Pi$ are the identity and permutation
matrices (\ie~~$I_{ab}^{a'b'}=\del_a^{a'}\del_b^{b'}$~ and
{}~$\Pi_{ab}^{a'b'}=\del_a^{b'}\del_b^{a'}$~ with $a,b,a',b'\in\{+,-\}$),
and
\eqn\So{ S_0(\th) ~=~
   {\Gamma({1\o 2}-{i\th\o 2\pi})~\Gamma(1+{i\th\o 2\pi}) \o
    \Gamma({1\o 2}+{i\th\o 2\pi})~\Gamma(1-{i\th\o 2\pi}) }
 ~=~\exp\Biggl\{i\int_0^\infty {d\om\o \om}
  {J_0(0) \sin(\om {2u\th\o \pi}) \o \cosh(\om u)}e^{-\om u}
  \Biggr\}~~.}
(Of course $J_0(0)=1$ on the most rhs of \So; it is inserted
there, as well as the arbitrary variable $u>0$ which can
simply be rescaled away, in order to exhibit an amusing
relation between the phase shift associated with $S_0(\th)$ and
$p_s(k)$ of \sdisp\ at $k$ such that ${\pi\sin k\o 2u}=\th$,
cf.~\kresc.)

\newsec{Discussion}

Eq.~\Sss\ is identified as the \sm\ of the massive sector of the
$SU(2)$-Thirring model [9-10,12-14].  
It is equal to the limit $g\to (-{\pi\o 2})^+$
of the \sm\ of the ordinary massive Thirring
model~\rzamsa\rWeisz~(which, up to a sign~\rher, is also that of the
sine-Gordon model in the limit $\beta\to \sqrt{8\pi}^-$).
Eq.~\Scc, on the other hand, has been recently
proposed~\rzamsb~(cf.~also~\rFT) to describe
the massless scattering theory associated with the level one $SU(2)$
WZW conformal field theory. {}Finally,
the fact that $S_{sc}(\th,\beta)$ turned out to
be rapidity-independent indicates that the massive and massless sectors
essentially decouple in the scaling limit.
These observations are all in concert with the identification of the
full scaled model as the $SU(2)$ chiral-invariant Thirring field theory.

Furthermore, as mentioned in
the introduction, this field theory is not quite a true tensor product of
two sub-theories, due to nontrivial ``gluing'' of sectors in their
spectra. In the spirit of~\rher, it appears that the
phase factor $i$ in \Ssc\ signals this effect at the
\sm\ level. It would be interesting to investigate what
consequences this factor may have on the correlation
functions of the theory.\foot{Recall the work of~\rBKW, where
the spin correlation functions in the Ising field theory~\rMTW~were
reconstructed using the form factor bootstrap program~\rSmir.
In this model, the nontrivial sign of the \sm\ $S(\th)=-1$
leads to correlators which are
expressed in terms of solutions to Painlev\`e equations~\rMTW,
rather than simple Bessel functions which arise in fermion
correlators in the trivial theory of a free massive Majorana fermion
whose \sm\ is simply $S(\th)=1$. Cf.~also~\rLash.}
However, for doing that a better understanding
of scattering theories involving  massless particles
is needed, as well as an extension
of the form factor bootstrap program to their framework.

\bigskip
\bigskip
{\it Acknowledgements.}
I would like to thank F.~E\ss ler for discussions.
This work was supported in part by the US-Israel Binational
Science Foundation.

\vfill\eject
\appendix{A}{Analysis of the dispersion relations}

We start with the massless  case which  is relatively simple.
To derive eq.~\cexp\ from \cdisp\ we first expand the Bessel functions
$J_\nu(\om)$ in powers of $\om$
and integrate the resulting series term by term,
using formulas 4.111(3,4,7) of \rGR.
This gives
\eqn\ca{ \eqalign{
 p_c(\lam) ~&=~ -{\rm sgn}~\lam~\Biggl\{ -{\pi\o 2}+2\arctan e^x \cr
 &~~~~~~~~~~+
  \sum_{k=1}^\infty {1\o 2^{2k} (k!)^2} \left({\pi\o 2U}\right)^{2k}
   \left(d\o dx\right)^{2k-1} {1\o \cosh x} \Biggr\}
    \Biggl|_{x=\pi|\lam|/2U}  \cr
 \eps_c(\lam) ~&=~
  \sum_{k=0}^\infty {1\o 2^{2k+1} k! (k+1)!} \left({\pi\o 2U}\right)^{2k+1}
   \left(d\o dx\right)^{2k} {1\o \cosh x}
    \Biggl|_{x=\pi|\lam|/2U} ~~. \cr} }
(Recall that throughout the paper $U>0$.)
Now for $x>0$ expand
$1/\cosh x=2\sum_{n=0}^\infty (-1)^n e^{-(2n+1)x}$
{}~and~ $2\arctan e^x=\pi-2\sum_{n=0}^\infty {(-1)^n \o 2n+1}e^{-(2n+1)x}$.
Interchanging summations over $n$ and $k$ and using the
power series expansion of $I_\nu(z)$, we obtain eq.~\cexp.

Since $I_\nu(z) \sim (2\pi z)^{-1/2} e^z$ for large $|z|$, we see that
the expansions \cexp\ absolutely converge for $|\lam|>1$, and in fact
they are convergent also for $|\lam|$=1. We note that a complementary
small-$(\lam/U)$ expansion can be obtained from \ca\ by expanding
$1/\cosh x=\sum_{n=0}^\infty {E_n\o n!}x^n$, where $E_n$ are
Euler's numbers. This way we arrive at
\eqn\cb{\eqalign{
 p_c(\lam)~&=~-\sum_{n=1 \atop n~{\rm odd}} {1\o n!}
   \left({\pi \lam \o 2U}\right)^n~\sum_{k=0}^\infty
   {E_{n+2k-1}\o 2^{2k} (k!)^2} \left({\pi\o 2U}\right)^{2k} \cr
 \eps_c(\lam)~&=~\sum_{n=0 \atop n~{\rm even}} {1\o n!}
   \left({\pi \lam \o 2U}\right)^n~\sum_{k=0}^\infty
   {E_{n+2k}\o 2^{2k+1} k!(k+1)!} \left({\pi\o 2U}\right)^{2k+1}~. \cr}}
This expansion converges for $|\lam|\leq U$.

\bigskip
In preparation to the analysis of the massive dispersion relation
\sdisp\ we need several definitions and lemmas.

\no {\bf Definition}: For $m=0,1,2,\ldots$
and integer $\ell$,  define the constants $A_\ell^{(m)}$
and  $B_\ell^{(m)}$ as the coefficients appearing in the expansions
\eqn\abdef{\eqalign{
 \left({d\o d\alp}\right)^{2m} {1\o \sqrt{\alp^2+1}} &=
  \sum_{\ell\in \ZZ} (-1)^{m+\ell} (4m-1-2\ell)!!~ \Alm
  (\alp^2+1)^{-(2m+{1\o 2}-\ell)}  \cr
 \left({d\o d\alp}\right)^{2(m+1)} \bigl(\sqrt{\alp^2+1}-\alp\bigr) &=
  \sum_{\ell\in \ZZ} (-1)^{m+\ell} (4m+1-2\ell)!!~ \Blm
  (\alp^2+1)^{-(2m+{3\o 2}-\ell)}  ~,\cr}}
where, as usual, the double-factorial stands for
$n!!=1\cdot 3\cdot\ldots \cdot n$ when $n$ is a positive odd integer and
$(-1)!!=1$.

The coefficients $\Alm,\Blm$ satisfy the following recursion relations
(for $m=1,2,\ldots$) and initial conditions:
\eqn\abrec{ \eqalign{
 \Alm = A_\ell^{(m-1)} +(4m-2\ell)A_{\ell-1}^{(m-1)}~~,~~~~&
 \Blm = B_\ell^{(m-1)} +(4m+2-2\ell)B_{\ell-1}^{(m-1)}~~\cr
 A_\ell^{(0)}~=~B_\ell^{(0)} ~&=~\del_{\ell,0}~~,\cr}}
from which it is easy to see that $\Alm=\Blm=0$ for
$\ell\not\in\{0,1,\ldots m\}$, so that the sums in \abdef\ are in fact
finite.

\no{\bf Lemma 1}: For $n=0,1,2,\ldots$
\eqn\lemi{ \eqalign{
 \sum_{\ell=0}^m (-1)^\ell (4m-1-2\ell)!!~ \Alm
  {2^n \Gamma(2m+n+{1\o 2}-\ell) \o \Gamma(2m+{1\o 2}-\ell)} &=
   {[(2m+2n-1)!!]^2 \o (2n-1)!!} \cr
 \sum_{\ell=0}^m (-1)^\ell (4m+1-2\ell)!! ~\Blm
  {2^n \Gamma(2m+n+{3\o 2}-\ell) \o \Gamma(2m+{3\o 2}-\ell)} &=
   {(2m+2n-1)!! (2m+2n+1)!! \o (2n-1)!!} ~.\cr }}

\no {\bf Proof}: Equate powers of $\alp^2$ on
both sides of \abdef\ after expanding them using
the binomial expansion $(\alp^2+1)^s=\sum_{n=0}^\infty
\pmatrix{s\cr n\cr} \alp^{2n}$, for $|\alp|<1$.  \qed

\medskip
\no {\bf Corollary}: For $k\in (-{\pi\o 2},{\pi\o 2})$
\eqn\cor{ \eqalign{
  k~&= \sum_{m=0}^\infty {\sin^{2m+1} k\o (2m+1)!}
    ~\sum_{\ell=0}^m (-1)^\ell (4m-1-2\ell)!!~ \Alm~\cr
  \cos k~&= 1-\sum_{m=1}^\infty {\sin^{2m} k\o (2m)!}
  ~\sum_{\ell=0}^{m-1} (-1)^\ell (4m-3-2\ell)!! ~B_\ell^{(m-1)}~.\cr}}

\no {\bf Proof}: For the first line use ~$k=\arcsin(\sin k)=
 \sum_{m=0}^\infty {[(2m-1)!!]^2\o (2m+1)!} \sin^{2m+1} k$~ and
then replace the numerator of the
coefficient here by the lhs of the first line of \lemi,
with $n$=0. For the second line expand ~$\cos k=\sqrt{1-\sin^2 k}=
1-\sum_{m=1}^\infty {(2m-3)!!~(2m-1)!!\o (2m)!}\sin^{2m} k$~ and
then use the second line of \lemi\ with $n$=0 and $m$ replaced
by $(m-1)$.    \qed

\medskip
\no {\bf Lemma 2}: For $m=0,1,2,\ldots$, let
\eqn\stdef{
  S_m(z)=\sum_{\ell=0}^m (-1)^\ell \Alm z^{2m-\ell}K_{2m-\ell}(z)~~,~~~~
  T_m(z)=\sum_{\ell=0}^m (-1)^\ell \Blm z^{2m+1-\ell}K_{2m+1-\ell}(z)~,}
where the $K_\nu(z)$ are modified Bessel functions. Then
\eqn\lemii{ S_m(z)~=~z^{2m}K_0(z)~~~,~~~~~~
            T_m(z)~=~z^{2m+1}K_1(z)~~.}

\no {\bf Proof}: Using \abrec\
and the recursion relation 8.486(10) of~\rGR~for
the  $K_\nu(z)$, one obtains
$S_{m+1}(z)=z^2 S_m(z)$ and $T_{m+1}(z)=z^2 T_m(z)$ with
$S_0(z)=K_0(z)$ and $T_0(z)=z K_1(z)$, from which \lemii\
immediately follows.  \qed

\medskip
Equipped with the above results, consider \sdisp.
Expanding the ~$\sin(\om \sin k)$~ and ~$\cos(\om \sin k)$~
in power series in ~$\om\sin k$~ and integrating term by term using
6.621(4)  of~\rGR~we find
\eqn\sa{\eqalign{
 p_s(k)&= k-2\sum_{m=0}^\infty {(-1)^m \o (2m+1)!}\sin^{2m+1} k
    ~\sum_{n=1}^\infty (-1)^{n-1} \left( {d\o d \alp}\right)^{2m}
    {1\o \sqrt{\alp^2+1}}\Biggl|_{\alp=2nU} ~\cr
 \eps_s(k)&=U-\cos k+2\sum_{m=0}^\infty {(-1)^m \o (2m)!}\sin^{2m} k
    ~\sum_{n=1}^\infty (-1)^{n-1} \left( {d\o d \alp}\right)^{2m}
  \bigl(\sqrt{\alp^2+1}-\alp\bigr) \Bigl|_{\alp=2nU} ~~.\cr}}
Now using \abdef\ and \cor\ yields, after reordering the summations
over $\ell$ and $n$,
\eqn\sb{\eqalign{
 p_s(k)~&=~ \sum_{m=0}^\infty {\sin^{2m+1} k\o (2m+1)!}
 ~\sum_{\ell=0}^m (-1)^\ell (4m-1-2\ell)!! ~\Alm \cr
  &~~~~~~~~~\times~\left(
  1+2\sum_{n=1}^\infty (-1)^n (\alp^2+1)^{-(2m+{1\o 2}-\ell)}
  \Bigl|_{\alp=2nU} \right) ~\cr
 \eps_s(k)~&=~U-1+2\sum_{n=1}^\infty (-1)^{n-1} (\sqrt{\alp^2+1}-\alp)
  \Bigl|_{\alp=2nU} ~\cr
 &+ ~~~~~~\sum_{m=1}^\infty {\sin^{2m} k\o (2m)!}
 ~\sum_{\ell=0}^{m-1} (-1)^\ell (4m-3-2\ell)!!~ B_\ell^{(m-1)}\cr
 &~~~~~~~~~~~~\times~ \left(
  1+2\sum_{n=1}^\infty (-1)^n (\alp^2+1)^{-(2m-{1\o 2}-\ell)}
  \Bigl|_{\alp=2nU} \right) ~~.\cr}}

The sums over $n$ encountered in the last step have been analyzed
in detail by Fisher and Barber in~\rFB, where they are referred to
as ``remnant functions'' (they are related to the so-called
Epstein-Hurwitz zeta function). Using the notation introduced in
eq.~(2.25) of~\rFB, we have
\eqn\rem{ \eqalign{
 \sum_{n=1}^\infty (-1)^n (\alp^2+1)^{\sig -1}\bigl|_{\alp=n/y}
 ~=~ &y^{2(1-\sig)} {\Gamma(\sig)\o \Gamma(-{1\o 2})}~
  R_{\sig,0}^{(-)}(y^2)~~~~~~~~(\sig= -{1\o 2},-{3\o 2},-{5\o 2},\ldots)\cr
 \sum_{n=1}^\infty {(-1)^{n} \o \sqrt{\alp^2+1}} \Biggl|_{\alp=n/y}
 ~&=~ -{y\o 2} \Bigl[ R_{1/2,0}^{(-)}(y^2)+2 \ln 2\Bigr]~~\cr
 \sum_{n=1}^\infty (-1)^{n-1}\bigl(\sqrt{\alp^2+1}-\alp\bigr)\Bigl|_{\alp=n/y}
 ~&=~ {1\o 4y} \Bigl[ R_{3/2,0}^{(-)}(y^2)+2y^2 \ln 2\Bigr]~~~.\cr}}
The crucial step now is to use eqs.~(2.26),
(6.20), and (6.21) of~\rFB, which leads to
\eqn\sc{\eqalign{
 p_s(k)~&= ~\sum_{m=0}^\infty {\sin^{2m+1} k\o (2m+1)!}
 ~\sum_{\ell=0}^m (-1)^\ell (4m-1-2\ell)!! ~\Alm \cr
  &~~~~~~~~~~~~~~~~~~~~~~~~\times ~{2U^{-1}\o (4m-1-2\ell)!!}
   ~\sum_{n=0}^\infty U_n^{2m-\ell} K_{2m-\ell}(U_n) \cr
  &=~ {2\o U} \sum_{m=0}^\infty {\sin^{2m+1} k\o (2m+1)!}
   ~~\sum_{n=0}^\infty S_m(U_n)~~,\cr
 \eps_s(k)~&=~{2\o U}\sum_{n=0}^\infty {K_1(U_n)\o U_n}
 ~+ ~\sum_{m=1}^\infty {\sin^{2m} k\o (2m)!}
 ~\sum_{\ell=0}^{m-1} (-1)^\ell (4m-3-2\ell)!! ~B_\ell^{(m-1)}\cr
  &~~~~~~~~~~~~~~~~~~~~~~~~~~~~~\times~{2U^{-1}\o (4m-3-2\ell)!!}
   ~\sum_{n=0}^\infty U_n^{2m-1-\ell} K_{2m-1-\ell}(U_n) \cr
 &=~  {2\o U}\sum_{n=0}^\infty {K_1(U_n)\o U_n}~+
   {2\o U} \sum_{m=1}^\infty {\sin^{2m} k\o (2m)!}
   ~\sum_{n=0}^\infty T_{m-1}(U_n)~~,\cr}}
where $U_n\equiv (n+{1\o 2}){\pi\o U}$, and in the second
lines in each formula we employ the definitions \stdef. Now
invoking \lemii\ and reordering the summations over $n$ and $m$,
we finally obtain \sexp.

\vfill\eject
\listrefs

\bye\end